# Improving photocatalytic activity of TiO$_2$ through reduction

Daoyu Zhang,[a] Minnan Yang,[b] Shuai Dong[a],*

Abstract: The rutile TiO$_2$ (110) surface reduced by the bridging oxygen vacancy, bridging hydroxyl group or Ti interstitial atom has been investigated by calculating their electronic structures using the density functional theory plus $U$ method. It is found that defect states located in the forbidden band gap can enhance optical absorption. When the surface is highly reduced, the defect states approach the valence band. More importantly, defects induce a substantial up-shift of the conduction band edge, rendering the reduced surface stronger reducibility. The shifts of both conduction and valence band edges are due to the dipole moments created by these defects.

One of the most versatile of technological and industrial materials is $TiO_2$, which is used as white pigment, gas sensors, varistors, optical coating, corrosion-protective coating, medical implant material, photocatalyst, and so on.[1] A majority of these mentioned applications involve the exposed facets of this material. Taking the rutile $TiO_2$ for example, the (110) surface is the most stable, widely used in applications, as well as a prototypical model system for scientific studies of $TiO_2$ on adsorption, dissociation, diffusion, and product formation.[2-4] Many elementary steps of the photocatalytic water decomposition, such as the electron transfer process, have been extensively investigated using the rutile $TiO_2$ (110) surface model.[5]

Usually, $TiO_2$ is slightly oxygen-deficient ($TiO_{2-x}$, $x\sim0.01$), namely, a reduced $n$-type semiconductor.[6] Moreover, when $TiO_2$ is annealed in ultra-high vacuum (UHV), it will lose some bridging oxygen atoms, giving rise to extra oxygen vacancies (O-vac's), and thus the concentration of oxygen deficiency may increase up to several percent. In addition, the chemically active oxygen vacancy can readily capture residual water even under the UHV conditions and create a pair of nearby bridging hydroxyl groups (O-H's).[7, 8] A plenty of O-H's at the $TiO_2$ (110) surface can also be introduced by the process of hydrogenation.[9, 10] Both the bridging oxygen vacancy and hydroxyl group give rise to the excess electrons, which will be trapped by specific Ti atoms.[11-14] In this sense, these Ti atoms are reduced with the nominal charge of +3, forming the defect states in the forbidden band. Besides the oxygen vacancy and hydroxyl group, another reductant defect can be the Ti interstitial (Ti-int),[15-18] which may occur during the sputter/annealing cycles in UHV.[19]

Reductant defects at the surfaces of a material can deeply influence many physical and chemical properties, e.g. the surface-to-adsorbate charge transfer, surface binding and reactivity due to the donating electrons.[20-25] For example, the presence of O-vac's, O-H's and Ti-int's at the rutile $TiO_2$ (110) surface can promote the capacity of $O_2$ adsorption, because the extra electrons from defects will transfer to the $O_2$'s $2p$ orbitals to increase the Coulombic attraction between anionic oxygen and Ti cations. Although extensive experimental and theoretical works have revealed the many effects of O-vac's, O-H's and Ti-int's at the rutile $TiO_2$ (110) surface like aforementioned oxygen absorption, little attention has been paid to the effects of these defects to the band gap and band edges, which are important quantities for photocatalysis. For semiconductors, the band gap determines the light absorption edge and the band edges determine the ability of transferring excited electrons or holes to the adsorbed species. Our previous work,[26] taking hydroxylation of the $TiO_2$(110) surface as an example, had shown that the reduced surface can change its energies in band edges towards to the vacuum level. The increased conduction band edge energy means that the hydroxylated $TiO_2$ surface enhances the reducing power for photocatalysis. In order to generalize these effects derived from hydroxylation, it is meaningful to study reduced surfaces resulting from other species of reductant defects such as O-vac and Ti-int. In this sense, this work aims to systematically discuss the effect of the reductant defects of O-vac and Ti-int, along with O-H, on the band gap, band edges, and the defect states of $TiO_2$ using the density functional theory (DFT) plus U calculations. The DFT+U method can reasonably describe the electronic structures when strongly correlated system like $TiO_2$ are considered. Previously, Morgan and Watson performed DFT+U calculations on the reduced $TiO_2$ and recovered the experimentally observed gap states well, in agreement with the prediction of the hybrid exchange functional (B3LYP).[13] Mulheran, *et al.* demonstrated DFT+U and atomistic charge equilibration calculations give accordant results for surface and interstitial Ti diffusion at the rutile $TiO_2$(110) surface.[17] Even though, influence of reductant defects at the $TiO_2$(110) surface on those photocatalysis-related quantities (e.g. the gap states, band gap, band edges) have not been systematically studied.

Our spin-polarized calculations were performed using the projector-augmented wave pseudopotentials as implemented in the Vienna *ab initio* Simulation Package (VASP).[27, 28] The electronic interactions are described within the LDA+U formalism. The Hubbard-type correction (*U*) was applied to Ti's 3*d* orbitals with the value of 5.5 eV which can give a proper description of the defect states in reduced $TiO_2$ according to previous studies.[29] The energy cutoff for plane wave basis was set to be 450 eV. The atomic positions were relaxed towards equilibrium using the conjugate gradient method until the force on each atom is less than 0.01 eV/Å. For calculating partial occupancies, Gaussian smearing with a width of 0.01 eV was employed.

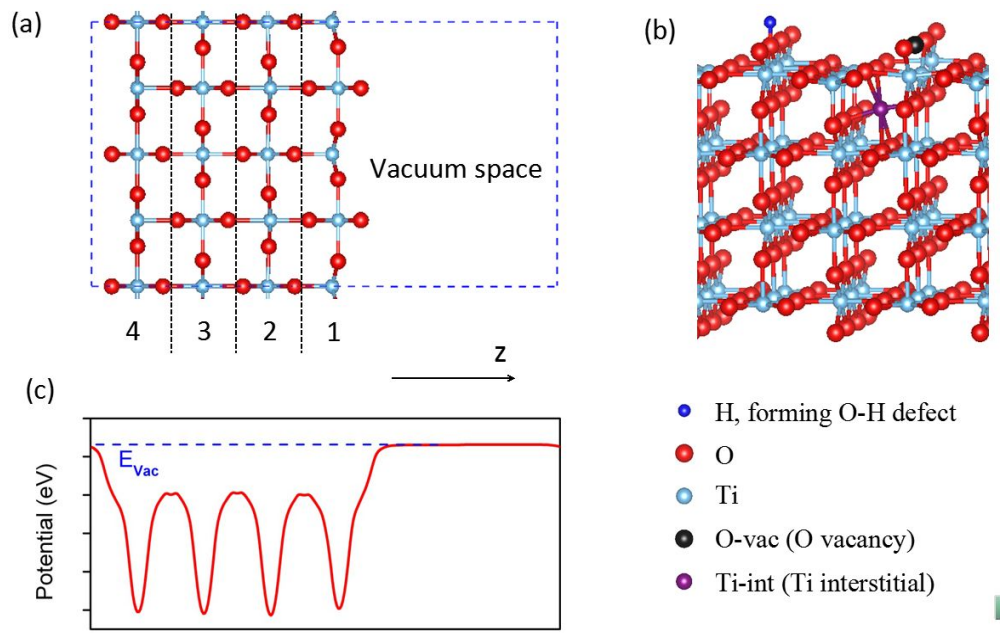

**Fig. 1** (Color online) (a) Model of the rutile TiO$_2$ (110) surface with four trilayers in which atoms in the third and fourth are fixed at their bulk positions. (b) Site indication of reductant defects used in this work. (c) Method for determining the vacuum level through the electrostatic potential of the lattice.

The experimental lattice parameters of $a$ = 4.593 Å and $c$ = 2.958 Å was used to construct the stoichiometric TiO$_2$ (110) surface. The surface was modeled by (3×2) periodically repeated slabs of four trilayers with a vacuum space of ~11 Å. Atoms of the third and fourth trilayers were fixed at their bulk positions as shown in Fig. 1(a). After the stoichiometric surface had been relaxed, the reduced surfaces were constructed: the reductant defect 'O-vac' was created by removing a single bridging O atom off; 'O-H' - by adding a H atom on a bridging O atom; 'Ti-int' - by inserting a Ti atom into an interstitial cavity in the first trilayer,[17] as schemed in Fig. 1 (b). Then these reduced slabs were used as the starting geometry for all the relaxations performed by DFT+U calculations. While $\Gamma$-point-only sampling was used for the geometrical relaxation of surfaces, automatically generated $\Gamma$-point-centered 3×2×1 Monkhorst-Pack mesh was used for static calculations. The monopole, dipole and quadrupole corrections have been applied to the electrostatic interaction between the slab and its periodic images in the direction perpendicular to the slab.

For the bands energy alignment, the vacuum energy level was selected as the common zero energy reference (shifted to 0 eV).[30-32] In our model system with surface, the vacuum level is determined by the (110)-planar average electrostatic potential in the vacuum region, where the electrostatic potential does not vary along the $z$ direction ($E_{vac}$ as shown in Fig. 1(c)). Then all eigenenergies subtracting $E_{vac}$ is the "absolute" energy values used in the following.

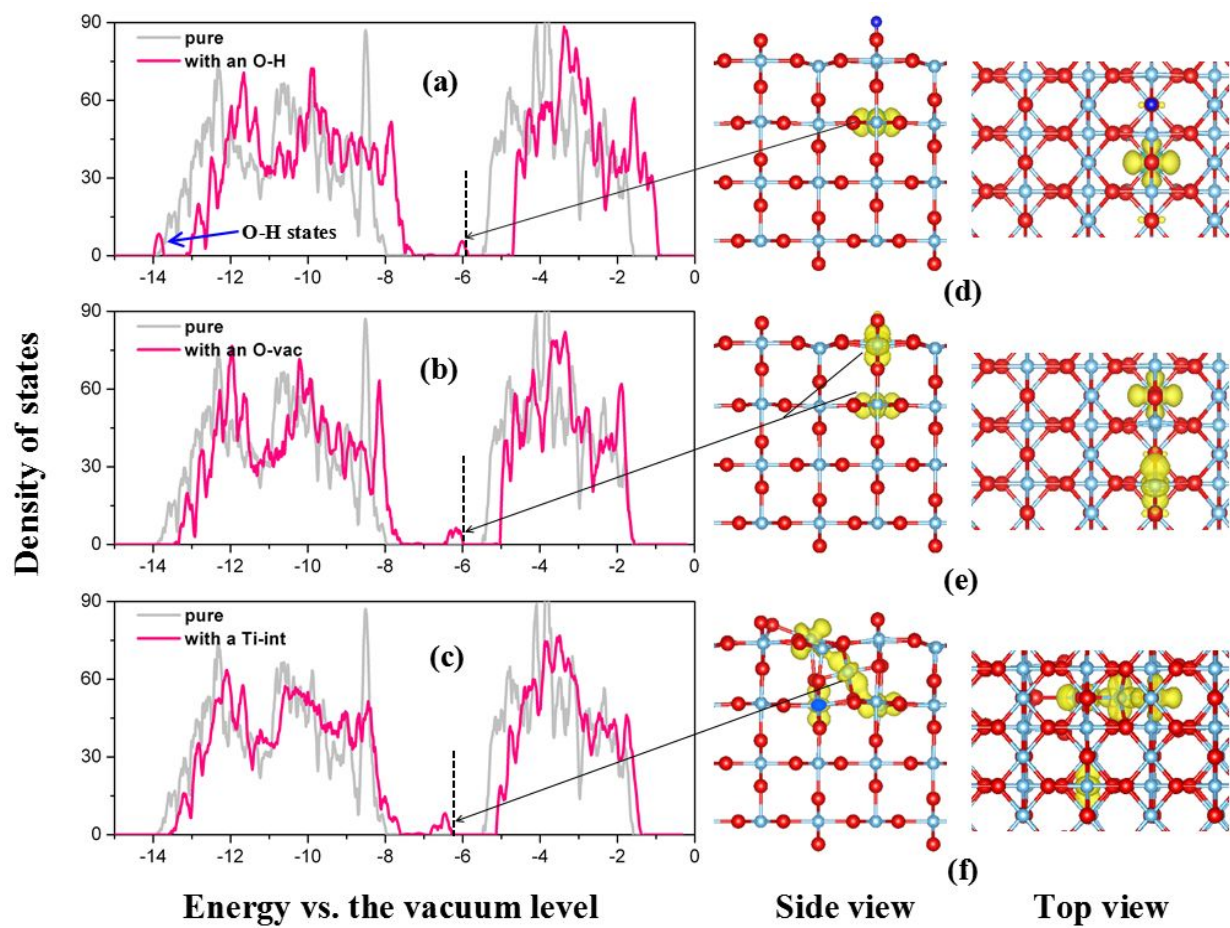

**Fig. 2** (Color online) The projected density of states (pDOS) on atoms in the first and second trilayers of the TiO$_2$ (110) surfaces with reductant defects, (a) O-H, (b) O-vac, and (c) Ti-int, in comparison with the pure surface (gray). The short, vertical, dash lines indicate the Fermi level. (d), (e), and (f) are the corresponding spin charge densities viewed from both side and top.

The electronic structures aligned with the vacuum level of the reduced TiO$_2$ (110) surfaces are calculated. Since the atoms of two bottom layers in the slab are fixed at their bulk positions, there is only one open surface (two top layers) to be relaxed, which projected density of states (pDOS) is shown in Fig. 2(a-c) for three reduced cases. In comparison with the unreduced surface, both the conduction band edge and valence band edge are shifted upward in energy despite the type of reduction. In addition, the defect states appear within the forbidden band. The changes of band gaps, band edges and the positions of defect states for reduced surfaces are summarized in Tab. I.

**Tab. I.** The calculated changes of the valence band edge, conduction band edge, and band gap ($\Delta E_V$, $\Delta E_C$, and $\Delta E_g$) of the surface with the bridging oxygen vacancy, bridging hydroxyl group, and Ti interstitial atom. Positive value indicates an increase in energy with respect to the pure surface. $p_z$ represents the dipole moment of the reduced surface and $\Delta V_z$ is defined by Eq. 1.

|  | defect states (eV) | $\Delta E_V$ (eV) | $\Delta E_C$ (eV) | $\Delta E_g$ (eV) | $p_z$ (Debye) | $\Delta V_z$ (eV) | $\varepsilon$ |
|---|---|---|---|---|---|---|---|
| O-H | -6.09 ~ -5.97 | 0.68 | 0.67 | -0.01 | 2.21 | 0.20[a] ~ 0.75[b] | 3.759 |
| O-vac | -6.37 ~ -6.08 | 0.38 | 0.36 | -0.02 | 1.26 | 0.11[a] ~ 0.42[b] | 3.783 |
| Ti-int | -6.76 ~ -6.38 | 0.33 | 0.36 | 0.03 | 1.30 | 0.12[a] ~ 0.43[b] | 3.749 |

[a]The value estimated by Eq. 1 where $\varepsilon$ is taken as the dielectric constant of the relevant reduced surface.

[b]The value estimated by Eq. 1 where $\varepsilon$ is taken as 1 (as in vacuum).

Based on these results, the reduced TiO$_2$ surface can exhibit enhancement of the photocatalytic activity apparently considering the following two aspects. First, a reduced surface can improve optical absorption of long wavelength photons via the defect states in the forbidden band. Occurrence of these defect states can effectively lower the electron transition energy and thus promote the creation of electron-hole pairs which act as reactants for next reduction-oxidation reactions. As shown in Fig. 2(d-f), the spin charge densities of defect states own the characteristics of *d* orbital at specific Ti sites apparently. These Ti's capture electrons transferred from the reductant defects, resulting in reduced $Ti^{3+}$. An O-H, O-vac or Ti-int reductant defect gives rise to one, two or four excess electrons to the surface respectively, and thus reduces one, two or four reduced Ti's.[33, 34]

As shown in Fig. 2(a-c), it is worthwhile to note that the width of defect states of reduced surfaces depend on the number of excess electrons. The more electrons donated by reductant defect to the lattice, the wider the defect states' bandwidth. Thus, the Ti-int defect generates the widest defect band among three reductant defects studied here. Besides, the highly reduced surface has lower defect states in energy, more close to the valence band edge. To further confirm these dependences, the electronic structures of the surfaces with different coverage of O-H defects have been calculated. For the surfaces covered by 1/3, 2/3, and 1 monolayer (ML) O-H, the widths of the defect states are of 0.38, 0.53, and 0.95 eV respectively, and the energy gaps between the valence band edge and the lower edge of defect band are 0.90, 0.84, and 0.67 eV, respectively. In other words, there is a trend of overlapping for the defect states and the valence band edge with increasing surface reduction. This tendency of overlapping is helpful to transfer the photoexcited holes to reactive sites at the catalyst surface.[35]

The above reduction-dependent defect band may give a reasonable explanation on the experimental observations of "black" $TiO_2$ nanoparticles prepared through hydrogenation of "white" $TiO_2$ in a 20.0-bar $H_2$ atmosphere.[9] The black $TiO_2$ exhibits substantial photocatalytic activity due to band tail states which are formed by the mid-gap states near the valence band edge. It is an open question where do the mid-gap states originate from. According to our calculations, the $Ti^{3+}$ cations reduced by the hydroxyl groups which cover surfaces of black $TiO_2$ nanoparticle to high extent, generate the band tail states.

The hydroxyl σ bonding O-H states, which locate at about -14 eV as seen from pDOS of the surface (Fig. 2(a)), do not contribute to the mid-gap states. And the O-H states are not possible for efficient hole traps, well in agreement with the previous prediction using B3LYP functional.[36]

Second, the substantial shifts in band edges of the reduced surface (see Tab. I) will influence the oxidizing power of holes in the valence band and the reducing power of electrons in the conduction band. Particularly, an increase in energy of the conduction band is very significant for the hydrogen revolution reaction during the photolysis process of water. It is well known that the conduction band edge (CBE) energy of rutile $TiO_2$ is only slightly greater than the electrochemical potential of $H^+/H_2$ redox couple.[37] Therefore, the up-shift of CBE due to occurrence of the kinetic overpotential is advantage to drive the hydrogen production reaction spontaneously by the excited electrons in the conduction band upon illumination. In fact, many efforts have been devoted to the up-shift of CBE.[38,39] Experiments have demonstrated that reducing of $TiO_2$ materials by hydrogenation is a simple and effective strategy to enhance the solar-to-hydrogen conversion efficiency of these materials.[9,40] The up-shift of the CBE energy revealed in our calculations can be an important origin of the improved photocatalytic activity.

Physically, the band-edges shifts of the reduced $TiO_2$ (110) surfaces can be attributed to the electric dipoles created by the reductant defects.[26] The stoichiometric rutile $TiO_2$ (110) surface itself is non-polar and has a neglectable dipole moment of only ~0.15 Debye given by our DFT+U calculation. However, when the surface is reduced, the dipoles are created. For example, the reductant defect O-H introduces a hydroxyl dipole and a polaronic dipole which locates at the site of the $Ti^{3+}$ ion.[26] The dipole moments of reduced surfaces along the direction perpendicular to the surface ($p_z$), are listed in the sixth column of Tab. I. The dipoles will generate a local electric field and thus the electrostatic potential will be modulated near the surface, as illustrated in Fig. 3. To estimate the variation of band edges $\Delta V_z$ induced by dipoles, the simple parallel-plate capacitor model is used,[41] and thus $\Delta V_z$ can be formulated as:

$$\Delta V_z = e \frac{p_z}{A \varepsilon \varepsilon_0}, \quad (1)$$

in which $A$ is the surface area and $\varepsilon$ is the effective dielectric constant of the surface layer. Because the chemical environment of the dipoles are different, for example, in the case of the surface with an O-H the hydroxyl dipole is exposed in vacuum and the polaronic one is embedded in the slab, the accurate decomposition of the total dipole moment into individual contribution is technically difficult. Even though, the approximate value range of $\Delta V_z$ can be estimated by considering the dielectric constants of the relevant reduced surface and vacuum, and the estimated values are listed in the last column of Tab. I. It is clear that the values of the band-edges shifts ($\Delta E_C$ and $\Delta E_V$) extracted from the DOS's of the reduced surfaces are just located within the estimated ranges of $\Delta V_z$.

The synchronous up-shift of $\Delta E_C$ and $\Delta E_V$ can be also naturally understood as the dipoles' effect. Since the electric field created by dipoles modulates electrostatic potential for every electron, both the valence and conduction band edges are equally shifted by such an electrostatic potential, which will result in a whole up-shift of DOS and unchanged band gaps of the reduced surfaces.

According to Tab. I, it is interesting that the variation of band edges is closer to the particular value of $\Delta V_z$ when $\varepsilon$ is taken as the dielectric constant of vacuum. In other words, in our DFT+$U$ calculations, the electric field created by dipoles should be mostly localized in the empty space near the dipoles. Our previous study also confirmed that the outmost hydroxyl contributes dominantly to the total dipole moment, much more than the embedded Ti polaron.[26]

Comparing with the O-vac, the Ti-int defect, which donates more electrons into the lattice, however, does not induce much more shifts of band edges. As shown in Fig.2(f), the dipoles created by Ti-int defect are mostly embedded under the surface and surrounding the defect. According to our previous study,[26] such embedded polarons are only weakly contribute to the total dipole moment comparing with the exposed hydroxyl, as well as the exposed O-vac here.

The exposed O-H defect can induce shifts of band edges efficiently, better than other two types of defects, due to its intrinsic dipole of the hydroxyl group. This argument can be further confirmed by comparing the surfaces with 1/3 and 2/3 ML hydroxyl coverages, which donate two and four electrons and also form two and four polarons (equal to an O-vac and a Ti-int respectively). $TiO_2$ (110) surfaces with 1/3 and 2/3 ML hydroxyl coverages have band-edges variations of ~1.08 and ~1.92 eV, and the corresponding dipole moments are 3.29 and 5.84 Debye, respectively. These values are much larger than the equal-electron-donor O-vac and Ti-int defects. Considering that the formation and properties of polarons are independent on the source of electrons,[34] one can infer that O-H's contribute to $p_z$ dominately. So chemisorption of polar radicle groups on surface is a good and simple strategy for tuning the band-edges energies to pursuit better performance of photocatalytic redox reactions.

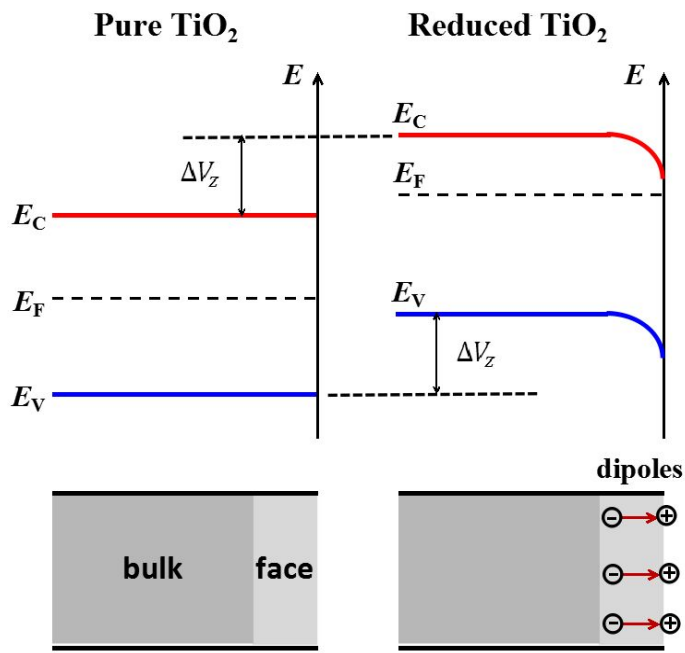

**Fig. 3** (Color online) Illustration of the dipoles at the surface created by a reductant defect. The electric field of dipoles causes the shifts in the band edges.

In order to demonstrate the effect of the value of U on the electronic properties studied here, the three reduced surface models created above used as initial geometric structures to repeat DFT+U calculations reported above with U values ranging from 3.0 to 5.5 eV. The band-gap variation of each of the reduced surface relative to the un-reduced surface is still small, less than 0.1 eV, for any U value employed in DFT+U calculations. For any U correction, the width of defect states in the forbidden gap becomes wider and defect states approach the valence band closer with increasing surface reduction. U values influence the band edge position weakly, and Fig.4 shows the dependence of $\Delta E_V$ on U. The dependence of $\Delta E_C$ is the same as $\Delta E_V$.

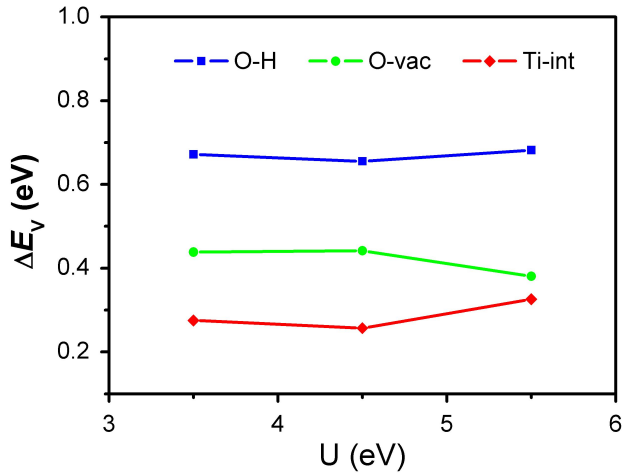

Fig. 4 (Color online) The weak dependence of $\Delta E_V$ on U correction. The dependence of $\Delta E_C$ is the same as $\Delta E_V$.

In summary, using the DFT+$U$ calculations, the effects of reductant defects, including hydroxyl group, oxygen vacancy, and Ti interstitial, have been studied on the $TiO_2$ (110) surface. It has been found that these defects do not change the band gap of the $TiO_2$ (110) surface. Even though, they can enhance the photocatalytic activity of $TiO_2$ from two aspects. One is the defect states occurring in the forbidden band of $TiO_2$, which can enhance the optical absorption. The other fact is the band-edges shift of both the conduction and valence bands to the vacuum level, which can effectively improve the reducing power of the electrons in the conduction band. Such a band-edges shift originates from the local field of electric dipoles created by defects. The mechanism of modulation of the band edges revealed here is general and applicable to other related materials.

## Acknowledgment

Work was supported by the NSFC (Nos. 51322206 and 11274060) and the 973 Projects of China (No. 2011CB922101).

## Notes and references

[a]Department of Physics, Southeast University, Nanjing, 211189, China.
[b]Department of Physics, China Pharmaceutical University, Nanjing, 211198, China
*Corresponding author. E-mail: sdong@seu.edu.cn